\documentclass[useAMS,usenatbib]{mn2e}
\usepackage{graphics,epsfig,lscape,txfonts,color}
\usepackage[english]{babel}
\usepackage{longtable}

\def\subsun{\mbox{$_{\odot}$}}
\def\ie{i.\,e.}                                      
\def\eg{e.\,g.}                                      

\def\gx{GX~339-4}

\title[]{Spectral properties of transitions between soft and hard state in GX~339-4}
\author[H. Stiele, S. Motta, T. Mu\~noz-Darias, and T. M. Belloni]{H. Stiele$^{1}$\thanks{E-mail:
holger.stiele@brera.inaf.it}, S. Motta$^{1,2}$, T. Mu\~noz-Darias$^{1}$, and T. M. Belloni$^{1}$\\
$^{1}$INAF-Osservatorio Astronomico di Brera, Via E. Bianchi 46, I-23807 Merate (LC), Italy \\
$^{2}$Universit\`a dell`Insubria, Via Valleggio 11, I-22100 Como, Italy}
\begin{document}

\date{2011 August 6}

\pagerange{\pageref{firstpage}--\pageref{lastpage}} \pubyear{2011}

\maketitle

\label{firstpage}

\begin{abstract}
We present a study of the spectral properties during state transition of \gx. Data are taken from the 2010 outburst of \gx, which is densely covered by Rossi X-ray Timing Explorer, providing an excellent coverage of the state transitions between the low/hard state and the high/soft state. We select all observations within a certain hardness ratio range during the soft intermediate state (SIMS). This sample was chosen in such a way to comprise all observations that show a type-B quasi-periodic oscillation (QPO). In addition, we also investigate the spectra of hard intermediate state observations. The spectra, obtained from Proportional Counter Array data in the 10 to 40 keV range, are fitted with a power law and an additional high energy cut-off if needed.  We find that the spectra are significantly harder during the SIMS of the soft-to-hard transition than they are during the hard-to-soft transition. This demonstrates that during the SIMS of the soft-to-hard transition not only the luminosity and peak frequencies of type-B QPOs are lower, but that also the photon index is lower, compared to the hard-to-soft transition. Hence, type-B QPOs can be associated to a different spectral shape even though they appear at the same hardness. However, in each branch only certain combinations of centroid frequency and photon index are realised. 
\end{abstract}

\begin{keywords}
X-rays: binaries -- X-rays: individual: GX~339-4 -- binaries: close -- black hole physics
\end{keywords}

\section{Introduction}
\gx, a black hole X-ray transient (BHT), was discovered in 1973 by the \textit{OSO-7} satellite \citep{1973ApJ...184L..67M}. Since then \gx\ showed several X-ray outburst \citep[see \eg\ ][]{1991ApJ...383..784M,1997A&A...322..857B,1997ApJ...479..926M,2004MNRAS.351..791Z,2005A&A...440..207B,2006MNRAS.367.1113B,2007ApJ...663.1309Y,2009MNRAS.400.1603M,2011MNRAS.410..679M}. The system is a low mass X-ray binary, harbouring a $>$6 M\subsun\ black hole accreting from a subgiant star in a  1.7 d orbital period \citep{2003ApJ...583L..95H,2008MNRAS.385.2205M}.  
\gx\ represents a prime example for the spectral and temporal evolution of a BHT during outburst. The different states through which a BHT evolves during an outburst can be identified in the hardness intensity diagram \citep[HID;][]{2005A&A...440..207B,2005Ap&SS.300..107H,2006MNRAS.370..837G,2006csxs.book..157M,2009MNRAS.396.1370F,2010LNP...794...53B}. In a log-log representation different states are found to correspond to different branches/areas of a q-shaped pattern.  Recently, \citet{2011MNRAS.410..679M} showed that the rms-intensity diagram (RID) allows the identification of states based on temporal information only. It is widely agreed on that at the begin and end of an outburst a BHT is in the so-called low/hard state (LHS), and that there is in between a transition to the high/soft state (HSS). In the HSS a strong thermal component, associated with disc emission is present, while spectra taken during the LHS show characteristics of hard (comptonized) emission. However, the exact definition of the states and especially of the transition between these states are still under debate. In this work we follow the classification of \citet{2010LNP...794...53B} \citep[see also][]{2005A&A...440..207B,2005Ap&SS.300..107H}, which comprises a hard as well as a soft intermediate state (HIMS/SIMS); see however \citet{2006csxs.book..157M} for an alternative classification and \citet{2009MNRAS.400.1603M} for a comparison. 

In this paper we investigate the spectral properties of transitions between the LHS and HSS. We focus on a comparison of the spectral properties of the SIMS observed during the hard-to-soft and the soft-to-hard transition. 

\section[]{Observations}
\begin{figure}
\resizebox{\hsize}{!}{\includegraphics[clip]{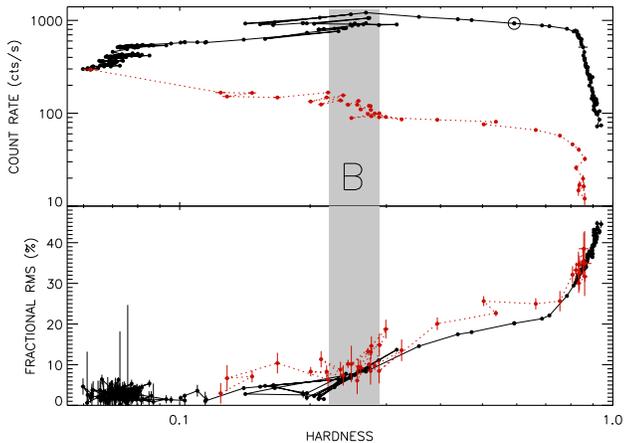}}
\caption{Upper panel: Hardness intensity diagram of the whole outburst, obtained using RXTE observations. Intensity corresponds to the count rate within the STD2 channels 0 -- 31 (2 -- 15 keV) and hardness is defined as the ratio of counts in 7 -- 13 (2.87 -- 5.71 keV) and 14 -- 23 (5.71 -- 9.51 keV) STD2 channels. Each point represents an entire observation. Consecutive observations are joined by a solid line until the softest observation is reached; after that by a (red) dashed line. The grey shaded area marks the hardness ratio range in which all observations that show type-B QPOs are located. This is also the hardness ratio range on which our study is based. Lower panel: corresponding hardness-rms diagram within the 0.1 -- 64 Hz frequency band. During the soft-to-hard transition the fractional rms is higher than during the hard-to-soft transition, but in the region selected for our study (grey shaded).}
\label{Fig:HID}
\end{figure}

The 2010 outburst of \gx\ was densely covered by Rossi X-ray Timing Explorer (RXTE), providing the (up to now) best coverage of the state transitions between the LHS and the HSS. For each observation of the outburst we determined the hardness ratio using Proportional Counter Unit 2 (PCU2) channels 7 -- 13 (2.87 -- 5.71 keV) for the soft band, and channels 14 -- 23 (5.71 -- 9.51 keV) for the hard band. The hardness intensity diagram (HID) as well as the hardness-rms diagram (HRD) of the whole outburst is shown in Fig.\,\ref{Fig:HID}. The fractional rms was computed within the 0.1 -- 64 Hz frequency band following \citet{1990A&A...227L..33B}. 

\gx\ shows the standard q-shaped pattern in the HID of this outburst. After the passage through the LHS and HIMS, type-B QPOs \citep[quasi periodic oscillations; for a description of different types of QPOs see][]{1999ApJ...526L..33W,2005ApJ...629..403C} are seen in the power density spectra (see Sect.\,\ref{Sect:PDS}), indicating that the system is in the SIMS. Once this state is reached, several fast transitions between the SIMS and the HSS are observed with one extending to the HIMS (see Fig.\,\ref{Fig:HID}). 
During the SIMS in the soft-to-hard transition at lower luminosities, the source also shows a few minor transitions. 

In the transition from the LHS to the HSS, in the following called the upper branch, we select all observations that show a type-B QPO. These observations have a hardness ratio ranging from 0.2208 to 0.2883 (grey shaded area in Fig.\,\ref{Fig:HID}). Please note that this hardness ratio range also includes observations without a type-B QPO. The same hardness ratio range is used to select observations from the lower branch (the back transition from the HSS to the LHS at lower luminosities). All observations showing type-B QPOs on either branch are within the hardness ratio range used. On the upper branch the centroid frequency of the type-B QPOs is $\sim$5 Hz, while on the lower branch it reduces to $\sim$2 Hz \citep[see][]{2011arXiv1108.0540M}. The fractional rms of observations showing type-B QPOs lies in the expected 5 -- 10 \% range \citep[see][]{2011MNRAS.410..679M}.  Furthermore, observations with a type-B QPO have a lower fractional rms during back transition than in the hard-to-soft transition. For most of the remaining parts of the outburst the fractional rms during back transition is higher than it has been in the hard-to-soft transition (observations belonging to the back transition are marked in red in Fig.\,\ref{Fig:HID}). 

In summary, observations were included in the sample for comparison if their hardness ratio ranged between 0.2208 and 0.2883, as this interval comprises all observations with type B-QPOs. We also analysed HIMS observations (0.2883 $<$ HR $\le$ 0.8) to cover the whole transition. 
 
\section[]{Power density spectra} 
\label{Sect:PDS}
We produced power density spectra (PDS) using 16 second long stretches of \textsc{GoodXenon}, \textsc{Event} and  \textsc{SingleBit} mode data. We limited PDS production to the Proportional Counter Array (PCA) channel band 0 -- 35 (2 -- 15 keV).  
After averaging the PDS and subtracting the contribution due to Poissonian noise \citep[see][]{1995ApJ...449..930Z}, the PDS were normalised according to \citet{1983ApJ...272..256L} and converted to square fractional rms \citep{1990A&A...227L..33B}. We fitted the PDS using the \textsc{Xspec} fitting package by applying a one-to-one energy-frequency conversion and a unit response. The noise components were fitted with three broad Lorentzian shapes, one zero-centered and other two centered at a few Hz \citep{2002ApJ...572..392B}. We fitted the QPOs with one Lorentzian each. Occasionally it was possible to achieve a significant improvement of the value of reduced $\chi^2$ by adding a Gaussian component, which better approximates the shape of the narrow peaks. The QPO centroid frequency as well as the total fractional rms for each observation is listed in Table~\ref{tab:data}.

\begin{figure}
\resizebox{\hsize}{!}{\includegraphics[clip]{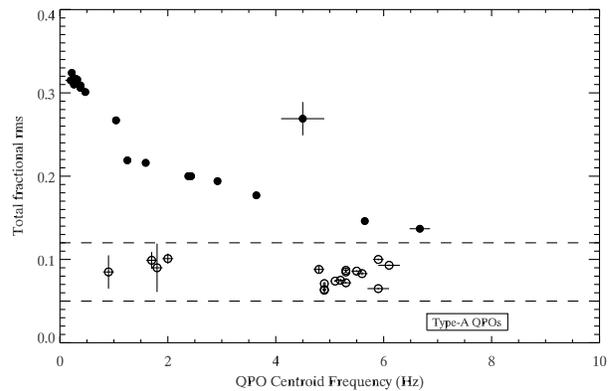}}
\caption{QPO centroid frequency vs. 0.1 -- 64 Hz fractional rms. Each point corresponds to a different observation. Symbols correspond to QPO types: filled circles are type-C QPOs, open circles are type-B QPOs. No type-A QPOs are detected in the 2010 outburst of \gx, but their location according to \citet{2011arXiv1108.0540M} is indicated.}
\label{Fig:QPO_cf_rms}
\end{figure}

\begin{figure*}
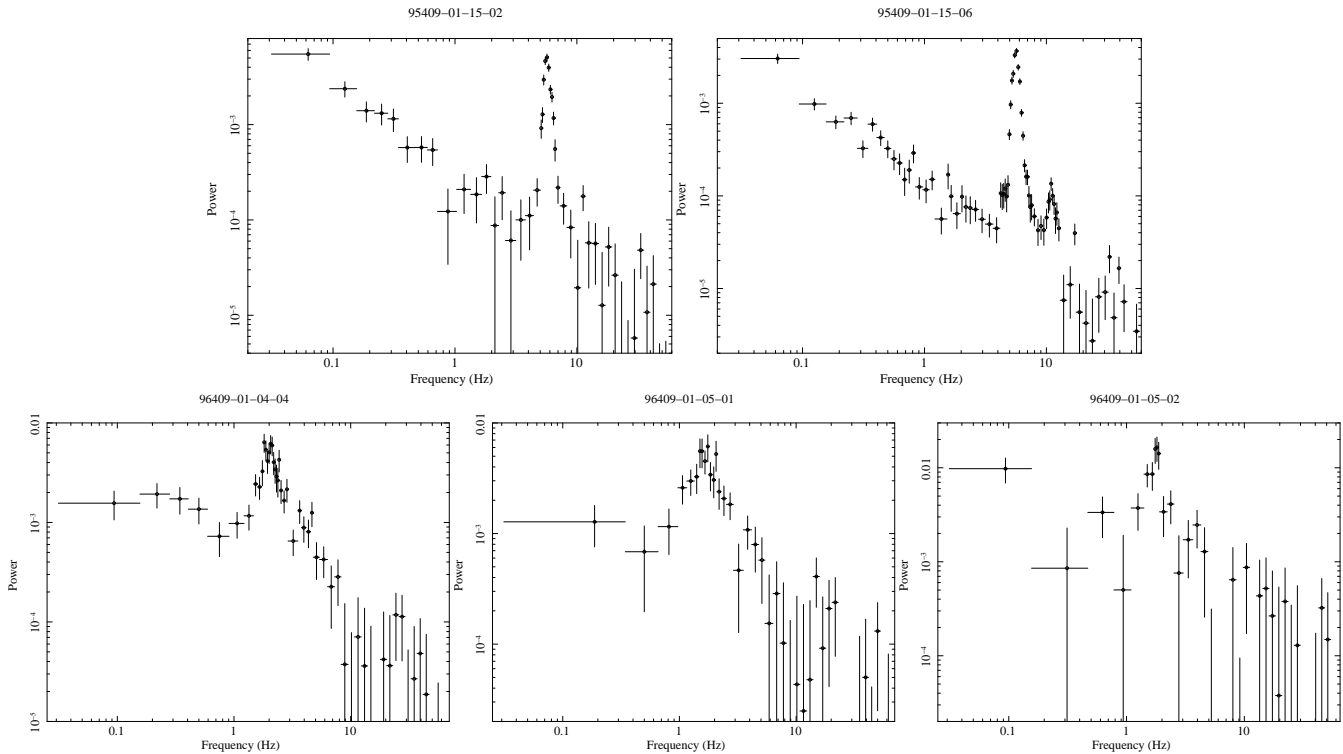

\resizebox{\hsize}{!}{\hskip10.0cm\includegraphics[clip, angle=-90]{pds_15_02_3.eps}\hskip0.2cm\includegraphics[clip, angle=-90]{pds_15_06_3.eps}\hskip10.0cm}
\resizebox{\hsize}{!}{\includegraphics[clip, angle=-90]{pds_04_04_3.eps}\hskip0.2cm\includegraphics[clip, angle=-90]{pds_05_01_3.eps}\hskip0.2cm\includegraphics[clip, angle=-90]{pds_05_02_3.eps}}
\caption{Power density spectra of some observations with type-B QPOs, detected during the hard-to-soft (upper row) and soft-to-hard transition (lower row).}
\label{Fig:PDS}
\end{figure*}

\citet{2004A&A...426..587C,2005ApJ...629..403C} showed that QPOs could be classified by the following properties:  the \emph{quality factor} ($Q=\nu_{centroid}/FWHM$) and the shape of the noise associated with the oscillation in the PDS. It has been also proven that QPOs can be discriminated by the amount of noise present in the PDS, which is quantified by the fractional rms \citep{2011MNRAS.410..679M}. This leads to a more specific version of the ABC classification \citep{2011arXiv1108.0540M}, in which type-B QPOs are characterised by rms strictly within the 5 -- 10\% interval and are observed at frequencies ranging from 1 to 7 Hz. 

The relation between centroid frequency and fractional rms is shown in Fig.\,\ref{Fig:QPO_cf_rms}. This plot allows us to distinguish clearly between type-C (filled circles) and type-B QPOs (open circles). Similar plots for XTE~J1859+226 and a sample of sources can be found in \citet{2004A&A...426..587C} and \citet{2005ApJ...629..403C}, respectively. The four points associated to type-B QPOs detected in the soft-to-hard transition (open circles with QPO centroid frequency between 0.9 and 2.0 Hz) are located in an area which is empty in the \citet{2004A&A...426..587C,2005ApJ...629..403C} plots, as these plots contain only type-B QPOs detected during the hard-to-soft transition. Furthermore, the four points associated to type-B QPOs during the soft-to-hard transition are located clearly outside the regions associated to type-C and type-A QPOs. Therefore we identify these four QPOs as type-B. A selection of PDS with type-B QPOs observed during the hard-to-soft as well as the soft-to-hard transition is shown in Fig.\,\ref{Fig:PDS}.  For a more detailed study of different types of QPOs detected in a sample comprising data of four outbursts of \gx\ and a discussion of QPO properties with respect to source flux as well as spectral properties see \citet{2011arXiv1108.0540M}. 
 
\section[]{Spectral analysis}
Energy spectra were extracted from PCA PCU2 data using the standard RXTE software within \textsc{heasoft} V.~6.9. We had to exclude all High Energy X-ray Timing Experiment (HEXTE) data from our analysis, since most of the HEXTE spectra contain strong residuals that are related to the difficulties in determining the background contribution in the spectra since the ``rocking" mechanism of HEXTE is broken.
To account for residual uncertainties in the instrument calibration a systematic error of 0.6 per cent was added to the PCA spectra\footnote{A detailed discussion on PCA calibration issues can be found at: http://www.universe.nasa.gov/xrays/programs/rxte/pca/doc/rmf/pcarmf-11.7/}. The spectra were fitted with \textsc{isis} V.~1.6.1 \citep{2000ASPC..216..591H}.

Since we were only interested in the behaviour of the hard spectral component we decided to neglect the contribution of the soft (disc) component and to focus our spectral analysis on the high energy range. 
We fitted the PCA (10 -- 40 keV) data using a power law model. For the first few observations of the HIMS an additional high energy cut-off was needed to obtain good fits. 

The fits resulted in formally acceptable reduced $\chi^2$ values and the spectra were clearly free of residuals. As an iron line was present at lower energies, indicating the presence of reflection processes \citep{1999MNRAS.303L..11Z}, it might be possible that also a broad reflection feature showed up around 30 keV \citep[\eg][]{1995MNRAS.273..837M}. This feature could be due either to reflection of the comptonized emission onto the accretion disk \citep[see][for a review]{2010LNP...794...17G} or to extended layers of material covering certain regions of the optically thick accretion disk \citep[e.g. wind clouds surrounding the central black hole, see][]{2006ApJ...643.1098S}. To exclude that the value of the photon index was affected by the presence of this feature we also tried a different model, consisting of a power law multiplied by a reflection component (\texttt{reflect}). We found that the relative reflection component was only needed for the first few HIMS observations. 

Taking reflection into account changes the value of the photon index in each individual observation. This has to be expected, since a ``phenomenological model'' such as a power law is not able to take into account the reflection features around 30 keV.
However the overall behaviour of the photon index observed between upper and lower branch stays the same (Fig.\,\ref{Fig:plotPI}). In the remaining observations the relative reflection component found was always consistent with zero within errors. In the following, we discuss only the results from the PCA 10 -- 40 keV spectra.

\section[]{Results and Discussion}
\begin{figure}
\resizebox{\hsize}{!}{\includegraphics[clip]{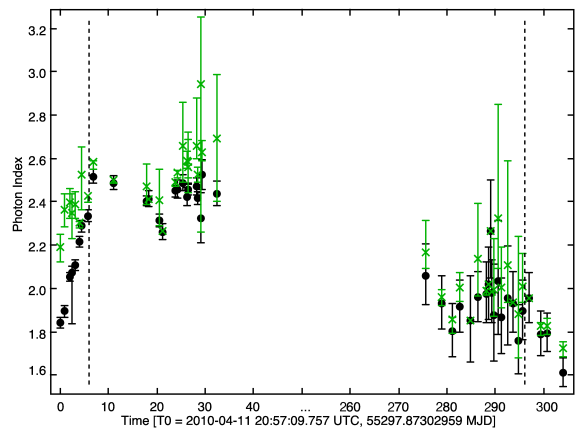}}
\caption{Temporal evolution of the photon index for different spectral models. The values obtained from a power law + cut-off model are given in black, the one obtained from a power law model with reflection in green. The dashed lines mark the first transition from HIMS to SIMS as well as the last transition from SIMS to HIMS.  Transitions in between are not indicated (see however Table~\ref{tab:data}). The time of the first observation of the HIMS in the hard-to-soft transition was selected as T$_0$.}
\label{Fig:plotPI}
\end{figure}

\begin{figure}
\resizebox{\hsize}{!}{\includegraphics[clip]{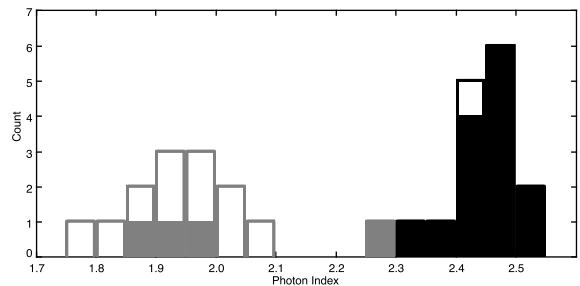}}
\caption{Distribution of the photon index values in the upper (black) and lower (grey) branch, with a bin size of 0.05.\@ Only the values obtained with the power law + cut-off model are shown. In the upper branch the photon index peaks at $\sim$2.45, while the distribution in the lower branch is broader and peaks at $\sim$1.95. Photon indices obtained from observations with type-B QPOs are marked as filled bars.}
\label{Fig:histPI}
\end{figure}

In spite of \gx\ has been deeply studied by RXTE during the last years, spectral transitions are very fast, which results in a few SIMS observations per outburst. Most of these observations are usually concentrated in the upper branch. Since the SIMS was intensively covered during 2010 and many type-B QPOs could be observed, it was possible to make an unprecedented  detailed direct comparison between the upper and lower branch SIMS. Hence,   
we investigated PCA spectra (10\,--\,40\,keV) to search for differences in the spectral properties of both branches and how they can be addressed in the light of (non)detection of radio emission and QPO properties. Figure~\ref{Fig:plotPI} shows the temporal evolution of the photon index. It increases during the HIMS and reaches a more or less constant value during the SIMS in the upper branch. In the lower branch, the SIMS spectra have a consistently lower photon index than those in the upper branch. The HIMS shows a moderate decrease in photon index. 
The fact that the soft-to-hard transition takes place at a much lower  photon index (see also Fig.\,\ref{Fig:histPI}) means that the high energy part of the spectrum is harder in the lower branch than it was in the upper branch. 
There are numerous works on BHTs that study the evolution of the temporal and spectral properties during the whole outburst \citep[\eg][]{2009ApJ...698.1398M,2010MNRAS.408.1796M,2010ApJ...723.1817S}. Other papers are dedicated to a detailed investigation of those properties during the outburst decay \citep[\eg][]{2004ApJ...603..231K,2005ApJ...622..508K,2006ApJ...639..340K}. The overall evolution of the photon index throughout a BHT outburst, which -- roughly speaking -- consists of an increase from the LHS to the HSS followed by a decrease when the source goes back to the LHS, is already known from those studies. However none of these studies focuses on the behaviour of the photon index during the upper and lower branch SIMS, which is most probably related to the previously sparsely sampling of (lower branch) SIMSs.

Remember that the observations were selected from the same hardness ratio range. A different contribution of the disc to the overall emission -- which arises in a reduced inner disc temperature -- compared to the upper branch has to be expected due to the lower luminosity in the soft-to-hard transition (\citealt{2003MNRAS.338..189M}; see \eg\ \citealt{2010ApJ...723.1817S} or \citealt{2011arXiv1103.4312S} for XTE~J1752-223 or \citealt{2011MNRAS.415..292M} for MAXI~J1659-152). This change in the disc contribution to the emission requires that also the rest of the spectrum must be modified since the hard tail depends (in a complicated way, via Comptonization processes; see \eg\ \citealt{1980A&A....86..121S,1995ApJ...450..876T} and references therein for thermal Comptonization, \citealt{1996A&AS..120C.171T} and references therein for non-thermal Comptonization, \citealt{2008MNRAS.390..227D} and references therein for hybrid Comptonization) from the disc itself. Thus the spectral shapes between upper and lower branch at a similar hardness ratio should differ from each other. While hardness ratio can be used as a good tracker of  the spectral shape for observations obtained at more or less the same flux, our results show that one should be cautious especially when analysing spectra at very different fluxes, as it is the case for observations of the upper and lower branch. In this case it is necessary to ``re-calibrate" its meaning according to the new flux level. In other words, the same hardness range tracks different photon index intervals at different fluxes.

\begin{figure}
\resizebox{\hsize}{!}{\includegraphics[clip]{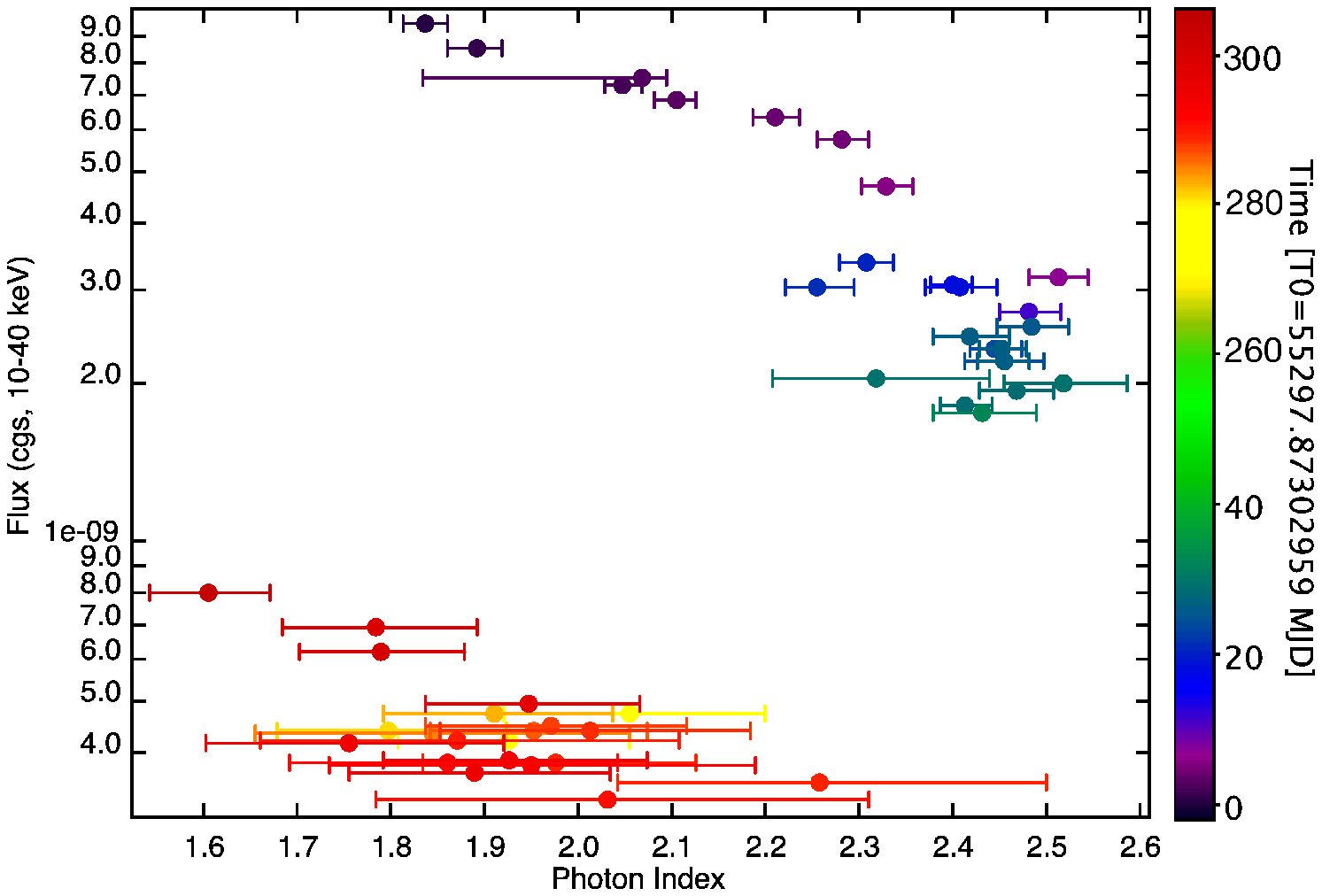}}
\resizebox{\hsize}{!}{\includegraphics[clip]{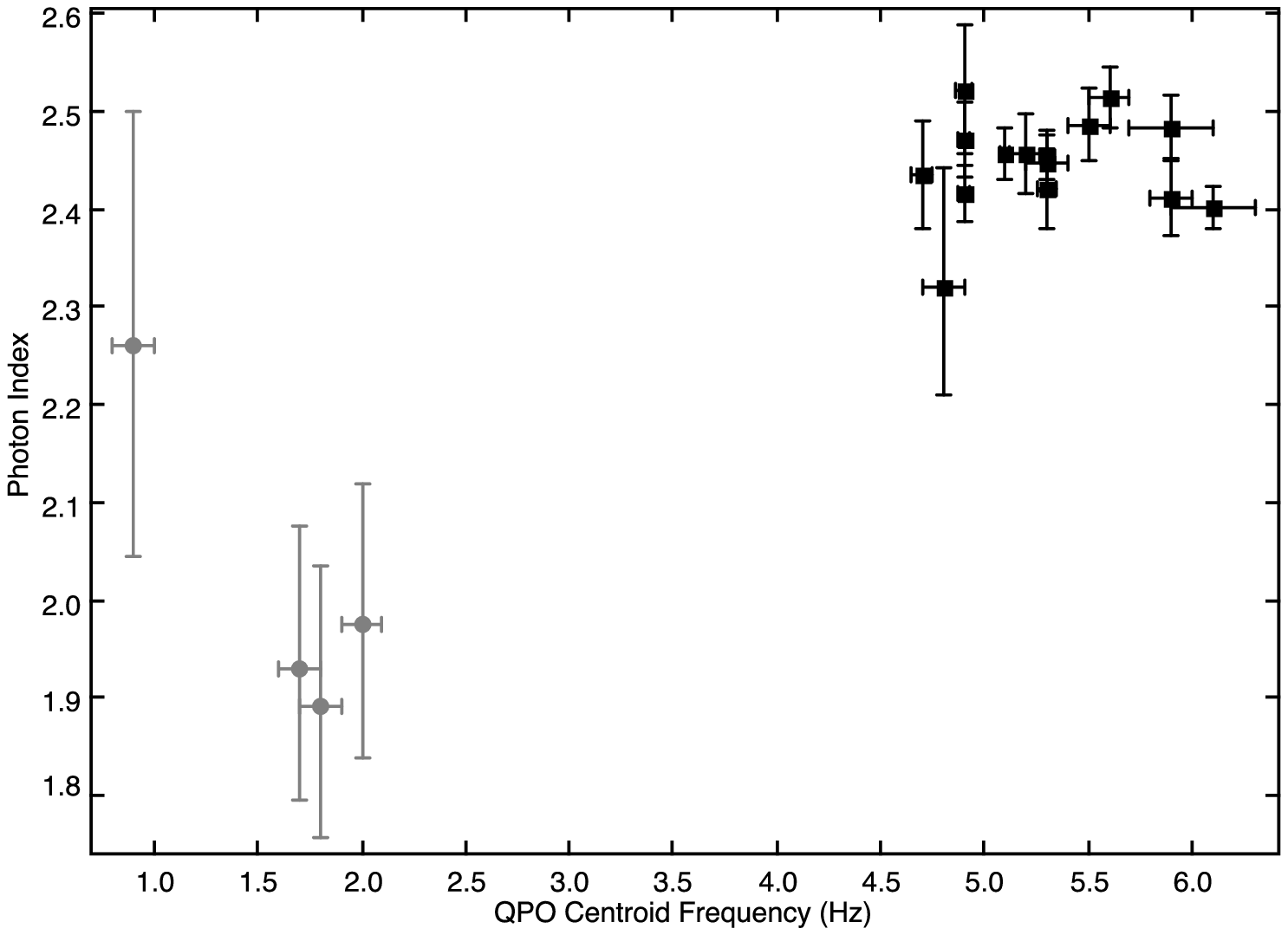}}
\caption{\textit{Upper pannel:} Relation between source flux and photon index. The colour of the dots represents the temporal evolution of the parameters during the outburst. \textit{Lower pannel:} Relation between QPO centroid frequency and photon index. Values belonging to observations of the hard-to-soft transition are marked as squares, those of the soft-to-hard transition as dots.}
\label{Fig:FluxPI}
\end{figure}

The connection between source flux and photon index is shown in Fig.\,\ref{Fig:FluxPI}. The colour of the dots represents the temporal evolution of the parameters during the outburst. The behaviour seen is expected according to the HID.     

The observations of the upper and lower branch were selected in such a way to include all observations with type-B QPOs and have a similar fractional rms.  
Furthermore, they are all located within a certain hardness ratio range, which is represented as a vertical strip in the HID. It is known that the jet line does not follow a vertical line in the HID, but shows a more complex behaviour \citep{2009MNRAS.396.1370F}. In the lower branch the jet does not turn on before the source is settled in the LHS \citep{2006ApJ...639..340K,2011ATel.3191....1R}. 
We showed that the photon index during the SIMS in the lower branch transition is smaller than in the upper branch transition. This means that lines of constant photon index run from the upper right to the lower left in the HID during state transition. 

It is known that the disappearance of the steady radio jet in the upper branch and the re-appearance of the jet in the lower branch take place at different hardness ratios. 
Our finding of a lower photon index during the SIMS of the lower branch implies that the jet appears at a much lower photon index in the soft-to-hard transition than it disappeared at in the upper branch. This implies that there is either no relation between photon index and radio emission or that this relation has to be extremely complicated. 
 
The photon indices found in the SIMS of the lower branch correspond to values obtained at the onset of the HIMS in the upper branch (see encircled dot in Fig.\,\ref{Fig:HID}).
This finding is in agreement with the lagging of timing properties compared to spectral properties in the soft-to-hard transition, as reported in \citet{2004ApJ...603..231K}.
Furthermore \citet{2011MNRAS.410..679M} showed that the onset of the upper branch HIMS and the transition form the lower branch HIMS to the LHS are situated on two different hard lines in the intensity-rms diagram of \gx. 

Furthermore, we found that type-B QPOs can be associated to different spectral shape. Previous studies, which did not differentiate between type-B and other types of QPOs, used a relation between photon-index and QPO centroid frequency to estimate the mass of the black hole in X-ray binaries \citep{2007ApJ...663..445S,2009ApJ...699..453S}.
In the upper branch type-B QPOs appear at a centroid frequency of $\sim$5 Hz \citep{2011arXiv1108.0540M} and in observations where the photon index is between 2.3 and 2.6.\@ In the lower branch the centroid frequency of type-B QPOs reduces to $\sim$2 Hz \citep{2011arXiv1108.0540M} and they are mainly observed in observations which spectra have a photon index of $\sim$1.9 (see Fig.\,\ref{Fig:FluxPI}).\@ However, we observe type-B QPOs neither in observations of the upper branch that have a photon index of $\sim$1.9 nor in observations of the lower branch which have a photon index of $\sim$2.4.\@ Therefore, the physical conditions that lead to the QPO do not depend in an obvious way on the power law parameters. Figure~\ref{Fig:histPI} clearly shows that within each branch the photon index is not enough to distinguish between observations with and without type-B QPOs in the selected sample. The difference is just in the timing. 
In addition, \citet{2011arXiv1108.0540M} showed that there is a correlation between centroid frequency of type-B QPOs and the count rate. We do not want to claim here a relation between centroid frequencies of type-B QPOs and photon index, but our findings underline that type-B QPOs can appear only in a very narrow range of properties of a BHT that are realised during state transitions. These properties occur in a rather narrow window in the HID and they are characterised by selected combinations of peak frequencies of type-B QPOs and photon index in the upper and lower branch, respectively. 

Recent works indicate that the hard component observed in spectra of intermediate states arises from hybrid Comptonization \citep[\eg][and references therein]{2008MNRAS.390..227D}. In the intermediate states, a hard tail extended to the MeV is observable in the spectra \citep[][like those observed in the HSS \citep{1998ApJ...500..899G}]{2007A&ARv..15....1D}. This intermediate state tail is softer than the tail that can be observed in the hard state and comes either in the form of a cut-off power law rolling over at energies larger than 150 keV \citep{2009MNRAS.400.1603M} or in the form of a simple power law, with no observable cut-off up to a few MeV \citep{2007ApJ...669..534C}. The extent of the tail shows that there must be non-thermal Compton scattering, as in the high/soft state. However, the tail is steeper, which means that the mean electron energy is lower. The observed spectral shape can only be produced by a combination of thermal/non-thermal Compton scattering. This could be produced in a single region filled with two populations of electrons (one thermal and the other non thermal). Alternatively there could be two different regions, one with thermal electrons (the same electrons that in the hard state are responsible of the thermal comptonization) and one with non-thermal, perhaps related to the jet base.
Assuming that the hard component results from hybrid Comptonization, the appearance of type-B QPOs depends directly on the properties of the hybrid Comptonizing medium. In particular, the type-B QPO frequency is strictly related to the temperature distribution of the electrons, that in turn determines the inclination of the hard part of the spectrum. In this regard, it appears natural to assume that the properties (\ie\ the temperature distribution) of such a medium are different in the softening and hardening phase, which are separated by the HSS, where the hard contribution to the emission is nearly negligible and therefore, our results are consistent with this scenario. 

However, the presence of diverse elements acting together in spectral/temporal evolution of the system (such as different populations of electrons, accretion flow and even a magnetic field) results in several effects that are difficult to disentangle.
The complexity of the situation makes it difficult to identify how the behavior of the emitting components could affect the properties of the QPOs and to determine whether the appearance of type-B QPOs is related to changes in the corona, in the disc, the accretion-ejection process, or a combination of several processes. Such issue is not trivial to the understanding of accretion and of the fundamental physics acting in presence of compact objects and strong gravitational fields and should be the topic of further investigations.
 
\section{Conclusion}
We investigated the spectral properties in the 10 -- 40 keV band during state transitions in the 2010 outburst of \gx. The sample of SIMS observations used contained all observations with type-B QPOs. Comparing the (mean) photon index found in the SIMS of the hard-to-soft transition with the one of the soft-to-hard transition a flatting of the power law is clearly evident. This means that the back transition from the soft to the hard state does not only occur at lower luminosity and with lower peak frequencies of type-B QPOs, but also at lower photon index. 
Hence, type-B QPOs can be associated to different spectral shape. However, in each branch only certain combinations of centroid frequency and photon index are realised.   

\section*{Acknowledgments}
The research leading to these results has received funding from the European Community's Seventh Framework Programme (FP7/2007-2013) under grant agreement number ITN 215212 ``Black Hole Universe". SM and TB acknowledge support from grant ASI-INAF I/009/10/0.

This work makes use of EURO-VO software, tools or services. The EURO-VO has been funded by the European Commission through contracts RI031675 (DCA) and 011892 (VO-TECH) under the 6th Framework Programme and contracts 212104 (AIDA) and 261541 (VO-ICE) under the 7th Framework Programme.

\bibliographystyle{mn2e}
\bibliography{/Users/apple/work/papers/my2010}

\appendix

\section[]{Online Material}
\begin{table*} 
\begin{minipage}{126mm}
\caption{Photon index and flux (10 -- 40 keV) derived from the best fit for each observation. A model consisting of a power-law, and -- if needed -- a cut-off at high energies was used.}
\begin{tabular}{llrcccccc}
\hline\noalign{\smallskip}
 \multicolumn{1}{c}{Obs. id.} &  \multicolumn{1}{c}{MJD} & \multicolumn{1}{c}{state} & \multicolumn{1}{c}{$\Gamma^{\dagger}$} &  \multicolumn{1}{c}{Flux$^{*}$}&  \multicolumn{1}{c}{fold energy}&  \multicolumn{1}{c}{cutoff energy} &  \multicolumn{1}{c}{QPO centroid} &  \multicolumn{1}{c}{total}\\
  & & & & &\multicolumn{1}{c}{[keV]} &\multicolumn{1}{c}{[keV]} &  \multicolumn{1}{c}{frequency} &  \multicolumn{1}{c}{fractional rms (\%)$^{\ddagger}$} \\
\hline\noalign{\smallskip} 
95409-01-14-02 & 55297.9 & HIMS & 1.84$\pm$0.02  & 9.43 & 132 $_{-34}^{+64}$ & 19.9 $_{-2.1}^{+2.6}$ & 1.25$\pm$0.01 & 21.9$\pm$0.1 \\
95409-01-14-03 & 55298.7 & HIMS & 1.89$\pm$0.03 & 8.47 & 65 $_{-22}^{+40}$ & 23.1 $_{-4.5}^{+3.5}$& 1.59$\pm$0.01 & 21.6$\pm$0.2 \\
95409-01-14-06 & 55299.8 & HIMS & 2.05$\pm$0.02 & 7.25 & 61 $_{-40}^{+68}$ & 26.2$_{-5.5}^{+6.3}$& 2.43$\pm$0.01 & 20.0$\pm$0.1  \\
95409-01-14-04 & 55300.3 & HIMS & 2.07$_{-0.24}^{+0.03}$ & 7.45 & 152 $_{-142}^{+187}$ & 20.6$_{-20.6}^{+39.7}$& 2.38$\pm$0.01 & 20.0$\pm$0.2  \\
95409-01-14-07 & 55300.9 & HIMS & 2.10$\pm$0.02 & 6.78 & 165 $_{-145}^{+86}$ & 18.45$_{-2.0}^{+3.8}$& 2.92$\pm$0.01 & 19.4$\pm$0.1  \\
95409-01-14-05 & 55301.8 & HIMS & 2.21$_{-0.02}^{+0.03}$  & 6.25 & -- & -- & 3.64$\pm$0.02 & 17.7$\pm$0.2 \\
95409-01-15-00 & 55302.2 & HIMS & 2.28$\pm$0.03 & 5.73 & -- & -- & -- & -- \\
95409-01-15-01 & 55303.6 & HIMS & 2.33$\pm$0.03 & 4.63 & -- & -- & 5.65$\pm$0.04 & 14.61$\pm$0.1 \\
95409-01-15-02 & 55304.7 & SIMS & 2.51$\pm$0.03 & 3.14 & -- & -- & 5.6$\pm$0.1 & 8.3$\pm$0.2 \\
95409-01-15-06 & 55309.0 & SIMS & 2.48$\pm$0.03 & 2.70 & -- & -- & 5.9$\pm$0.2 & 6.5$\pm$0.1 \\
95409-01-16-05 & 55315.7 & SIMS & 2.40$\pm$0.02 & 3.02 & -- & -- & 6.1$\pm$0.2 & 9.3$\pm$0.1 \\
95409-01-17-00 & 55316.1 & SIMS & 2.41$\pm$0.04 & 3.00 & -- & -- & 5.9$\pm$0.1 & 10.0$\pm$0.1 \\
95409-01-17-02 & 55318.4 & HIMS & 2.31$\pm$0.03 & 3.34 & -- & -- & 6.67$\pm$0.19 & 13.69$\pm$0.13 \\
95409-01-17-03 & 55319.1 & HIMS & 2.26$\pm$0.04 & 3.00 & -- & -- & -- & -- \\
95409-01-17-05 & 55321.7 & SIMS & 2.44$\pm$0.03 & 2.30 & -- & -- & 5.3$\pm$0.1 & 7.2$\pm$0.1 \\
95409-01-17-06 & 55322.2 & SIMS & 2.45$\pm$0.04 & 2.18 & -- & -- & 5.2$\pm$0.1 & 7.5$\pm$0.2 \\
95409-01-18-00 & 55323.2 & SIMS & 2.48$\pm$0.04 & 2.53 & -- & -- & 5.5$\pm$0.1 & 8.6$\pm$0.1 \\
95335-01-01-07 & 55324.2 & SIMS & 2.42$\pm$0.04 & 2.41 & -- & -- & 5.3$\pm$0.04 & 8.7$\pm$0.2 \\
95335-01-01-00 & 55324.3 & SIMS & 2.45$\pm$0.02 & 2.30 & -- & -- & 5.3$\pm$0.03 & 8.5$\pm$0.1 \\
95335-01-01-01 & 55324.4 & SIMS & 2.45$\pm$0.03 & 2.17 & -- & -- & 5.1$\pm$0.02 & 7.4$\pm$0.1 \\
95335-01-01-05 & 55326.2 & SIMS & 2.47$\pm$0.04 & 1.91 & -- & -- & 4.9$\pm$0.03 & 7.1$\pm$0.2 \\
95335-01-01-06 & 55326.3 & SIMS & 2.41$\pm$0.03 & 1.79 & -- & --& 4.9$\pm$0.03 & 6.3$\pm$0.2 \\
95409-01-18-04 & 55327.0 & SIMS & 2.32$_{-0.11}^{+0.12}$ & 2.01  & -- &-- & 4.8$\pm$0.1 & 8.8$\pm$0.4 \\
95409-01-18-05 & 55327.3 & SIMS & 2.52$\pm$0.07 & 1.96 & -- & -- & 4.9$\pm$0.04 & 6.4$\pm$0.4 \\
95409-01-19-00 & 55330.3 & SIMS & 2.43$_{-0.05}^{+0.06}$ & 1.73 & -- & -- & 4.7$\pm$0.05 & 3.3$\pm$1.1 \\
96409-01-02-02 & 55573.5 & SIMS & 2.05$_{-0.13}^{+0.14}$ & 0.47 & -- & -- & -- & -- \\
96409-01-03-00 & 55576.8 & SIMS & 1.93$_{-0.12}^{+0.13}$ & 0.42 & -- & -- & -- & -- \\
96409-01-03-01 & 55578.9 & SIMS & 1.80$_{-0.12}^{+0.13}$ & 0.43 & -- & --& -- & -- \\
96409-01-03-02 & 55580.6 & SIMS & 1.91$_{-0.12}^{+0.13}$ & 0.47 & -- & -- & -- & -- \\
96409-01-04-00 & 55582.7 & SIMS & 1.85$_{-0.19}^{+0.21}$ & 0.43 & -- & -- & -- & -- \\
96409-01-04-01 & 55584.4 & SIMS & 1.95$_{-0.11}^{+0.12}$ & 0.44 & -- & -- & -- & -- \\
96409-01-04-04 & 55585.9 & SIMS & 1.97$_{-0.13}^{+0.15}$ & 0.45 & -- & --& 2.0$\pm$0.1 & 10.1$\pm$0.6 \\
96409-01-04-02 & 55586.5 & SIMS & 2.01$_{-0.16}^{+0.17}$ & 0.43 & -- & -- & -- & -- \\
96409-01-04-05 & 55586.9 & SIMS & 2.26$_{-0.21}^{+0.24}$ & 0.35 & -- & -- & 0.9$\pm$0.1 & 8.5$\pm$2.0 \\
96409-01-04-03 & 55587.2 & SIMS & 1.98$_{-0.14}^{+0.15}$ & 0.38 & -- & -- & -- & -- \\
96409-01-04-07 & 55587.5 & HIMS & 1.87$_{-0.21}^{+0.24}$ & 0.42 & -- & -- & -- & -- \\
96409-01-04-08 & 55588.5 & SIMS & 2.03$_{-0.25}^{+0.28}$ & 0.32 & -- & -- & -- & -- \\
96409-01-05-00 & 55589.2 & SIMS & 1.86$_{-0.17}^{+0.18}$ & 0.38 & -- & -- & -- & -- \\
96409-01-05-04 & 55590.4 & HIMS & 1.95$_{-0.21}^{+0.24}$ & 0.37 & -- & -- & -- & -- \\
96409-01-05-01 & 55591.6 & SIMS & 1.93$_{-0.13}^{+0.15}$ & 0.39 & -- & --& 1.7$\pm$0.1 & 9.9$\pm$1.0 \\
96409-01-05-05 & 55592.7 & HIMS & 1.76$_{-0.15}^{+0.17}$ & 0.41 & -- & -- & -- & -- \\
96409-01-05-02 & 55593.5 & SIMS & 1.89$_{-0.13}^{+0.14}$ & 0.36 & -- & -- & 1.8$\pm$0.1 & 9.0$\pm$2.9 \\
96409-01-05-03 & 55594.9 & HIMS & 1.95$_{-0.11}^{+0.12}$ & 0.49 & -- & -- & -- & -- \\
96409-01-06-00 & 55597.3 & HIMS & 1.78$_{-0.10}^{+0.11}$ & 0.69 & -- & --& -- & -- \\
96409-01-06-01 & 55598.7 & HIMS & 1.79$\pm$0.09 & 0.62 & -- & -- & -- & -- \\
96409-01-06-02 & 55601.9 & HIMS & 1.61$_{0.06}^{+0.07}$ & 0.79 & -- & -- & -- & -- \\
\hline\noalign{\smallskip}
\end{tabular} 
~\\
Notes:\\
$^{\dagger}$: photon index\\
$^{*}$: flux in the 10 -- 40 keV band in units of 10$^{-9}$\,erg\,s$^{-1}$\,cm$^{-2}$\\
$^{\dagger}$: in the 0.1 -- 64 Hz range
\end{minipage} 
\label{tab:data}
\end{table*}

\bsp

\label{lastpage}

\end{document}